\documentclass[aps, superscriptaddress, prl, preprint]{revtex4-1}

\usepackage{graphicx}
\usepackage{dcolumn}
\usepackage{bm}

\begin{document}


\title{Spatial evolution of the ferromagnetic phase transition in an exchange graded film} 




\author{B. J. Kirby}
\email{bkirby@nist.gov}
\affiliation{Center for Neutron Research, NIST, Gaithersburg, MD 20899, USA}

\author{H. F. Belliveau}
\affiliation{Department of Physics, University of South Florida, Tampa, FL 33620}

\author{D. D. Belyea}
\affiliation{Department of Physics, University of South Florida, Tampa, FL 33620}

\author{P. A. Kienzle}
\affiliation{Center for Neutron Research, NIST, Gaithersburg, MD 20899, USA}

\author{A. J. Grutter}
\affiliation{Center for Neutron Research, NIST, Gaithersburg, MD 20899, USA}

\author{P. Riego}
\affiliation{CIC nanoGUNE Consolider, E-20018 Donostia - San Sebastian, Spain}

\author{A. Berger}
\email{a.berger@nanogune.eu}
\affiliation{CIC nanoGUNE Consolider, E-20018 Donostia - San Sebastian, Spain}

\author{Casey W. Miller}
\email{cwmsch@rit.edu}
\affiliation{School of Chemistry and Materials Science, Rochester Institute of Technology, Rochester, NY, 14623}


\date{\today}

\begin{abstract}
A combination of experiments and numerical modeling was used to study the spatial evolution of the ferromagnetic phase transition in a thin film engineered to have a smooth gradient in exchange strength.  Mean-field simulations predict, and experiments confirm that a 100 nm Ni$_{x}$Cu$_{1-x}$ alloy film with Ni concentration that varies by 9 $\%$ as a function of depth behaves predominantly as if comprised of a continuum of uncoupled ferromagnetic layers with continuously varying Curie temperatures. A mobile boundary separating ordered and disordered regions emerges as temperature is increased. We demonstrate continuous control of the boundary position with temperature, and reversible control of the magnetically ordered sample volume with magnetic field.   
\end{abstract}

\maketitle 
The precise fabrication of magnetic heterostructures can lead to control of the physical properties of the system, including magnetic ordering.  Ramos et al.\cite{Ramos_PRL_1990}, for example, showed that heterostructures composed of antiferromagnetic materials with nominally independent order parameters can exhibit a single phase transition at intermediate temperatures; Wang and Mills provided a mean-field theoretical treatment for such systems.\cite{Wang_PRB_1992}  Similarly, Marcellini et al., used finite size effects to study the novel magnetic ordering of Fe/V multilayers in which each layer had distinct ordering temperatures.\cite{Marcellini_PRB_2009}  Complementing these traditional heterostructures, recent work has explored materials with novel functionality derived from smoothly changing the films' physical properties during growth.   Examples applicable to next generation magnetic recording involved graded magnetic anisotropy,\cite{Suess_APL_2006,Berger_APL_2008,berger2010magnetic,Kirby_PRB_2010, Kirby_JAP_2015}  and the recent demonstration of a moveable antiferromagnet-ferromagnet phase boundary in doped FeRh films.\cite{LeGraet_APLmat_2015} 

This Letter considers a ferromagnetic thin film with exchange strength $J$ that varies continuously through its thickness. Within the mean field approximation, the exchange strength is proportional to the Curie temperature, meaning that a structure comprised of distributions of $J$ can be thought of as having a distribution of ``local" Curie temperatures, $T_{C}'$.  From a thermodynamics viewpoint, an exchange coupled multilayer is considered to have a single ``global" Curie temperature, $T_{C}$, corresponding to long-range order throughout the structure.\cite{Wang_PRB_1992,Skomski_JAP_2000}  However, the  degree to which exchange coupling impacts depth-dependent properties in real materials has not been thoroughly studied. We chose nickel-copper alloy as a model system to study this issue because it forms isomorphous solid solutions, and its $T_{C}$ changes linearly with composition.\cite{Ahern_PRS_1958,Kravets_PRB_2012}  A rational distribution of $J$ (and thus $T_{C}'$) can be made in films with well controlled depth dependent composition, e.g.,  Ni$_{x(z)}$Cu$_{1-x(z)}$, by varying the nickel content during growth. Mean-field simulations predict that the collective effect of exchange coupling in such systems is highly localized. Correspondingly, the ferromagnetic phase transition should proceed as expected from a set of uncoupled layers with distinct $J$. We verify these predictions with temperature and field dependent magnetization depth profiles determined from polarized neutron reflectometry (PNR).  Thus, we demonstrate the ability to control the displacement of a quasi phase boundary between effectively ferromagnetic and effectively paramagnetic regions in graded films using temperature and magnetic field.  This may have implications for a variety of thermomagnetic or spin caloritronic applications.  

\begin{figure}
\includegraphics{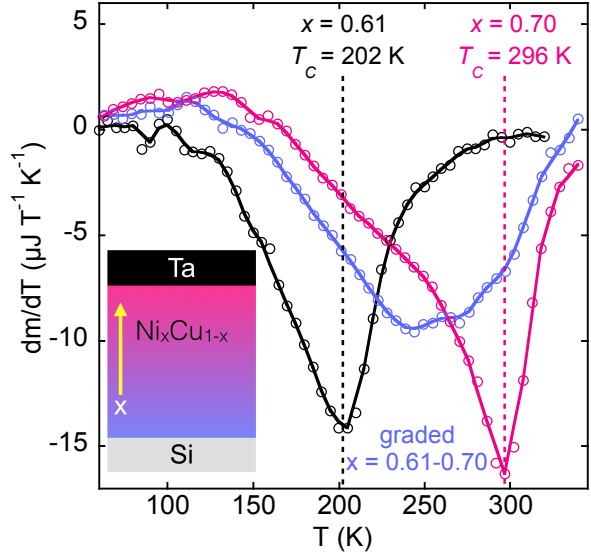}
\caption{\label{fig1}  Derivative of magnetic moment with respect to temperature for the uniform $x$ = 0.61 (black), graded $x$ = 0.61 -- 0.70 (purple), and uniform $x$ = 0.70 (pink) samples.}
\end{figure}
100 nm thick (111) textured Ni$_{x(z)}$Cu$_{1-x(z)}$ alloy films capped with 5 nm of Ta were deposited on Si substrates using room temperature sputtering.  The composition gradients were achieved by adjusting the deposition rates from two independent magnetron guns.  Computer control allowed the net deposition rate to remain constant throughout the co-deposition process.  The graded sample composition varied linearly from $x$ = 0.61 at the substrate interface to 0.70 at the top of the film (depicted in the Fig.~\ref{fig1} inset). Uniform control samples with $x$ = 0.61 and $x$ = 0.70 corresponding to the minimum and maximum compositions were also grown.  Figure~\ref{fig1} shows the derivative of the magnetic moment with respect to temperature ($T$) for the samples, as measured in a 60 mT field with vibrating sample magnetometry (VSM).  The $T$-dependent derivatives for the uniform samples exhibit distinct minima at 202 K and 296 K, indicative of their respective $T_{C}$.  The corresponding peak for the graded sample falls between those of the uniform samples, but is significantly broader for the graded film.  This suggests that the magnetic ordering takes place over a wide range of temperatures.  Given the intentionally engineered composition gradient, this smearing suggests that the ferromagnetically ordered fraction of the film shrinks in volume with increasing temperature, with that change propagating from the low toward the high Ni-content region.  Thus, it is interesting to consider how the boundary between the magnetically ordered and disordered regions moves through the structure in response to temperature and magnetic field. 

\begin{figure}
\includegraphics[scale=1]{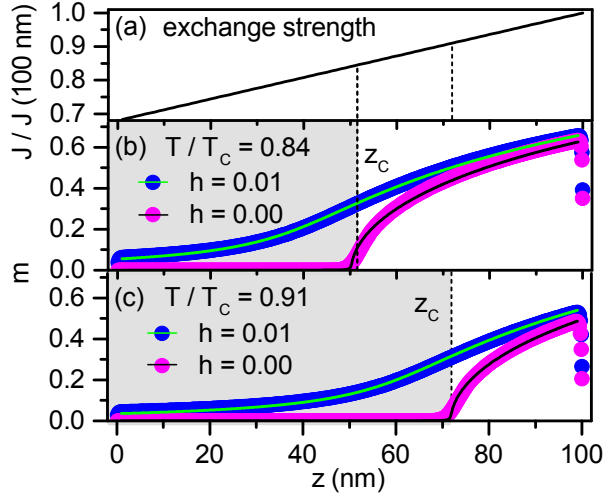}
\caption{\label{fig2}  Exchange strength gradient model.  (a) Depth-dependent exchange strength.  Note that $J$ is proportional to $T_{C}'$.  (b-c) Magnetization profiles at 0.84 $T_{C}$ and 0.91 $T_{C}$, respectively.  Symbols in (b-c) are the results of mean-field simulations, solid curves are fits to mean-field results using an closed-form function that neglects interlayer exchange coupling.  Dashed vertical lines in (b-c) indicate $z_{C}$, the boundary between the weakly (shaded) and strongly ordered regions.}
\end{figure}

The temperature and field dependent magnetic depth profile of the Ni$_{x(z)}$Cu$_{1-x(z)}$ film was treated theoretically via an exchange strength gradient model (ESG), which was solved numerically in mean field approximation (MFA).\cite{Berger_JMMM_1997,Riego_PRE_2015}  500 layers of ferromagnetically coupled spins were arranged on a fcc(111) lattice with 0.2 nm lattice spacing and depth dependent exchange coupling constant $J$. Given the translational invariance within the plane of each layer $i$, the spin variables in $i$ are resulting in one thermodynamic average for the magnetization $m_{i}$ only, a fact that implicitly introduces a quasi-infinite interaction range within each plane via the MFA.  This is in contrast to other methodologies, e.g. nearest-neighbor Monte Carlo simulations.\cite{Binder_PRB_1974}   However, the translational invaraiance is broken along the $z$ axis perpendicular to the (111) planes to account for the expected gradient in $J$, and thus $T_{C}'$ (Fig.~\ref{fig2} (a)).  It is important to note that the interlayer exchange coupling in the simulation is strong: half the total exchange field originates from interlayer coupling for the fcc(111) lattice. 

Figure~\ref{fig2} shows the results of several simulations in both zero ($h$ = 0.00) and nonzero ($h$ = 0.01) applied field for different temperatures (reduced to the maximum $T_{C}'$). The applied magnetic field $h$ is given as a fraction of the $T$ = 0 K exchange field, 6$JS$, and is thus unitless.  $m$ is also unitless, normalized by the depth-independent saturation magnetization.   When the temperature exceeds some of the $T_{C}'$ values with $h$ = 0.00 (Fig.~\ref{fig2} (b), lower curve), the layer divides into  strongly and weakly (effectively zero) ordered regions.  The boundary in space separating these regions, which we call the ``Curie depth", $z_{C}$, corresponds directly to the location with $T_{C}'$ equaling the sample temperature.  Once the applied field is increased to $h$ = 0.01 (upper curve, Fig.~\ref{fig2}(b)), the magnetization increases everywhere, particularly so near $z_{C}$.  As such, the boundary at $z_{C}$ can be thought of as separating a strongly ferromagnetic region from a region that exhibits paramagnetic character.  Fig.~\ref{fig2}(c) shows that $z_{C}$ increases significantly as $T$ increases from 0.84 to 0.91, moving closer to the high $T_{C}'$ end of the structure.  Thus, the simulation predicts a quasi phase boundary that can be moved continuously along the growth axis with temperature, and significantly modified with applied field.  Notably, both of these functionalities are reversible, as this is a second order phase transition without metastable states.   

Despite the inherently strong interlayer coupling in our model system, our numerical results appear to be very similar to what one would expect from magnetically decoupled layers with distinct $T_{C}$.  To investigate and eventually utilize this fact for our data analysis, we calculated the expected magnetization profiles for such uncoupled layer systems. In MFA, for a layer $i$ of  $S$ = $\frac{1}{2}$ spins with uniform exchange coupling strength $J$ and thus one specific local Curie temperature $T_{Ci}'$, the self-consistency equation to determine its temperature dependent magnetization $m_{i}$ is given by
\begin{equation}
m_{i}\left(T\right) = \tanh\left[\frac{Tc'_{i}}{T}\left(m_{i}\left(T\right)+h\right)\right].
\end{equation}

Dropping the subscript $i$ and defining the local Curie temperature as $T_{C}'\left(z\right) = T + a\left(z-z_{C}\right)$, Eq. (1) can be solved formally as
\begin{equation}
z = z_{C} + \frac{1}{A}\left[\frac{\tanh^{-1}\left(m\right)}{m+h} - 1\right],
\end{equation}
where $A$ = ${a}/{T}$ is the temperature normalized $T_{C}'$ gradient in units of inverse length.  The $z\left(m\right)$ function can be transformed into a closed form $m\left(z\right)$ expression, using a series expansion of tanh($x$) up to third order in $x$.\cite{supplemental}  This $m\left(z\right)$ function was fitted to the simulation data of the exchange gradient model using $z_{C}$ and $h$ as free parameters and is shown as solid curves in Fig. 2(b-c).  The agreement between the full simulation and the local $T_{C}'$ calculation is excellent, with the fitted values of $h$ matching the simulation inputs. Deviations only occur for vanishing $h$, and only within about 1 nm of $z_{C}$. This consistency suggests that interlayer exchange coupling does not lead to a large spatial spread of the phase transition in exchange graded films, and that the predominant temperature and field evolution of such films can be given by a continuum of $J$ or $T_{C}'$.  This is consistent with earlier results on 3-dimensional multiphase nanostructures, for which a coupling or penetration length of only a few interatomic distances was found.\cite{Skomski_JAP_2000}

To experimentally probe the nature of the ferromagnetic phase transition, polarized neutron reflectometry was used to determine the temperature and applied magnetic field ($H$) dependent magnetization depth profiles of the Ni$_{x(z)}$Cu$_{1-x(z)}$ sample.  Measurements were taken over a range of temperatures with in-plane fields of either 5 mT or 500 mT on the PBR beamline at the NIST Center for Neutron Research.\cite{supplemental}  For a PNR measurement,\cite{Chuck_book} model fitting of the $R^{++}$ (incident and scattered neutron moment parallel to $H$) and $R^{--}$ (incident and scattered antiparallel) are used to determine the depth profiles of the nuclear scattering length density $\rho_{N}\:=\:\Sigma_{i}N_{i}b_{i}$ (where $N$ is the number density, $b$ is the isotope-characteristic scattering length, and the summation $i$ is over each type of isotope being scattered from), and the component of the sample magnetization ($M$) parallel to $H$. 

PNR data measured in 500 mT at 145 K and 293 K are shown in Figure 3(a).  
\begin{figure}
\includegraphics[scale=1]{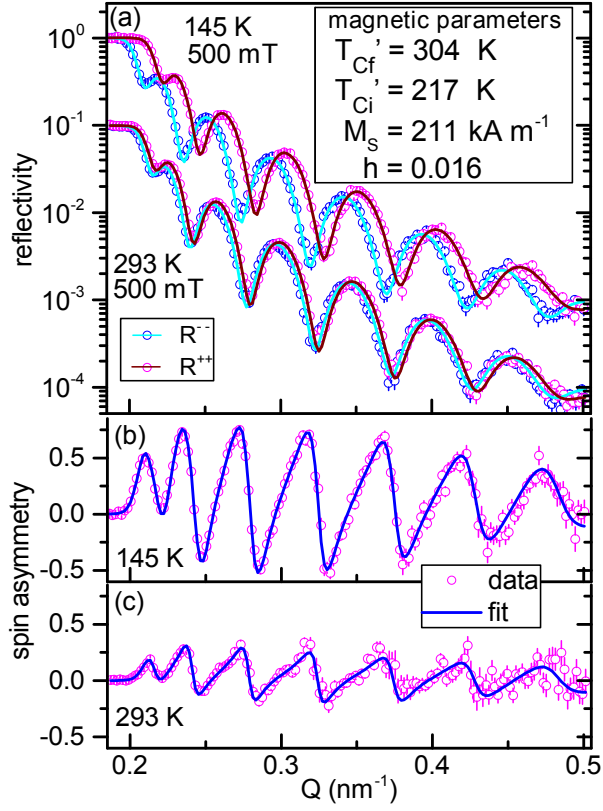}
\caption{\label{fig3}(a)  Fitted $R^{--}$ and $R^{++}$ at 145 K (top) and 293 K (offset by a factor of 10 for clarity) in 500 mT.  Inset shows magnetic parameters corresponding to the best-fit.  2$\sigma$ fitting uncertainties are $<$ 3 K for $T_{Ci}$ and $T_{Cf}$, $<$ 2 kA m$^{-1}$ for $M_{S}$, and $<$ 0.002 for $h$.  (b-c) Fitted data in (a) plotted as spin asymmetry.   Error bars on data correspond to $\pm$ 1 standard deviation.}
\end{figure}  
The clear spin-dependent oscillations demonstrate sensitivity to the nuclear and magnetic depth profiles.  To highlight the magnetic contribution to the scattering, the data in (a) are plotted in Fig.~\ref{fig3}(b-c) as spin asymmetry (the difference in $R^{++}$ and $R^{--}$ divided by the sum).  The amplitude of the spin asymmetry drops appreciably with increased $T$, corresponding to the drop in magnetization.    

At low $T$, the data can be fitted reasonably well by a model featuring a gradient in $M$, or even a NiCu layer of uniform composition and magnetization.  However, these models become progressively less effective as $T$ is increased.\cite{supplemental}  On the contrary, the 500 mT data at all $T$ are fit extremely well by the ESG model outlined above.   Specifically, the magnetic profiles are based upon the closed form $m\left(z\right)$, convoluted with a sloped line to account for the expected variation in total magnetic moment, and multiplied by the saturation magnetization ($M_{S}$) of the high $x$ end
\begin{equation}
M\left(z\right) = M_{S}\left(\frac{1-\frac{x_{i}}{x_{f}}}{t}z+\frac{x_{i}}{x_{f}}\right)m\left(z\right),
\end{equation}
where $t$ is the total thickness of the Ni$_{x\left(z\right)}$Cu$_{1-x\left(z\right)}$ layer, $x_{i}$ = 0.61, and $x_{f}$ = 0.70. We assume a linear distribution of $T_{C}'$, such that the $T$-dependent Curie depth can be described in terms of the top ($f$) and bottom ($i$) $T_{C}'$ \footnote{We note that lifting this constraint, and allowing $z_{C}$ to vary freely at each temperature does not significantly improve the fit quality, and yields $T$-dependent profiles very similar to those shown in Fig. 4(b).}   
\begin{equation}
z_{C}\left(T\right) = \frac{T-T_{Cf}'}{T_{Cf}'- T_{Ci}'}t.
\end{equation}
The data measured at all conditions (including 5 mT data discussed below) were simultaneously fit using a consistent model with only 5 magnetic parameters: $T_{Cf}'$, $T_{Ci}'$, $M_{S}$, $h\left(\mathrm{500\:mT}\right)$, and $h\left(\mathrm{5\: mT}\right)$.  
The fits are excellent, with examples shown as solid curves in Fig. 3, associated magnetic fitting parameters shown in the Fig. 3(a) inset, and the corresponding profiles shown in Figure 4 (a-b).  
\begin{figure}
\includegraphics[scale=1]{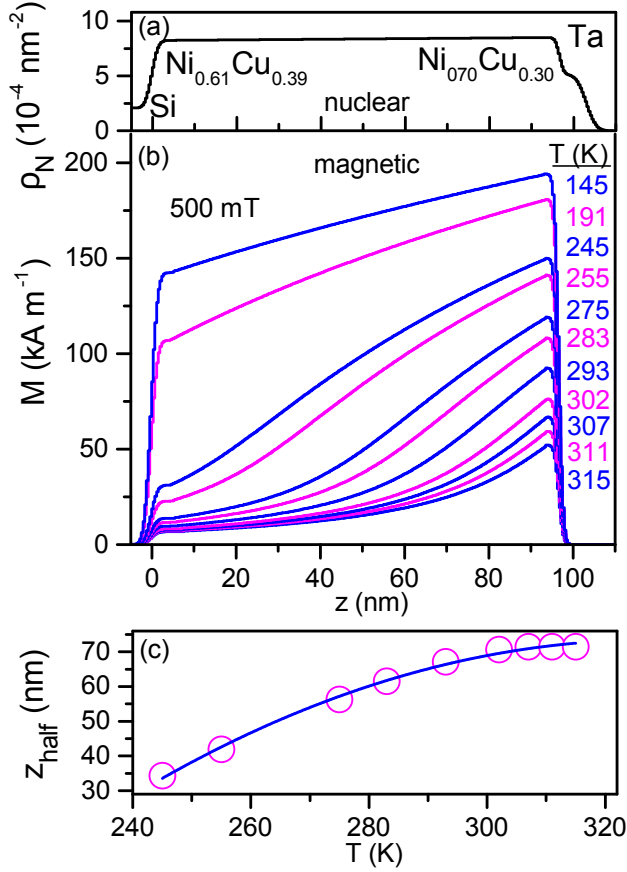}
\caption{\label{fig3}Nuclear (a) and magnetic (b) profiles corresponding to the fits in Fig. 3. (c) Temperature dependence of the depth corresponding to half the maximum magnetization.  Solid curve is a guide to the eye.}
\end{figure}  
The bottom and top values of $T_{C}'$ are 217 K and 304 K, with 2$\sigma$ fitting uncertainty less than $\pm$ 3 K.  Notably, this difference in $T_{C}'$ is very similar to the difference in $T_{C}$ estimated for the uniform boundary condition samples.\footnote{For the modeling, $T_{C}'$ corresponds to zero $m$ in zero $H$, while $T_{C}$ for the boundary condition samples was defined as the minimum in $\frac{dm}{dT}$ in finite $H$. Thus the absolute disagreement among the values is largely semantic.}   This $z$-dependent $T_{C}'$ is manifested in the Fig. 4(b) magnetic profiles.  At  145 K (well below $T_{C}$ of $x$ = 0.61 boundary condition sample), the entire Ni$_{x(z)}$Cu$_{1-x(z)}$ layer is strongly ordered.  But as $T$ exceeds 200 K, the depth profile becomes progressively more asymmetric, separating into weakly and strongly magnetized regions, with a boundary that moves continuously with $T$.  As $z_{C}$ is less obvious in high field, this moving boundary is highlighted in Fig. 4(c) in terms of $z_{half}$, the depth where the magnetization drops to half the maximum value.  Below 240 K, this condition does not exist anywhere in the Ni$_{x(z)}$Cu$_{1-x(z)}$layer, but from 240 K - 302 K $z_{half}$ moves monotonically towards the high $x$ end before leveling off as $T$ exceeds $T_{Cf}'$.

The ESG model also predicts that the quasi-phase boundary can be altered significantly with a large enough change in $H$ at a given $T$.  This is confirmed by the PNR data. At 275 K, $H$ was cycled between 5 mT and 500 mT multiple times, with PNR measurements performed at each step.  The amplitude of the spin asymmetry (Fig.~5(a-b)) drops for low fields, indicating a change in magnetic profile. The fits to the data are again quite good, with the two field conditions differing by only one fitting parameter, $h$, which is found to drop by almost an order of magnitude, from 0.016 to 0.003 (2$\sigma$ uncertainty less than 0.002). The corresponding profiles are displayed in Fig. 5(c).  At 5 mT, the sample is strongly magnetized near the top, with a magnetization near zero at the bottom.  Increasing the field to 500 mT appreciably magnetizes the region near the bottom, thereby drastically increasing the magnetized thickness of the film.   These data are cycle-independent, meaning that the quasi-phase boundary can be moved reversibly with magnetic field.

\begin{figure}
\includegraphics[scale=1]{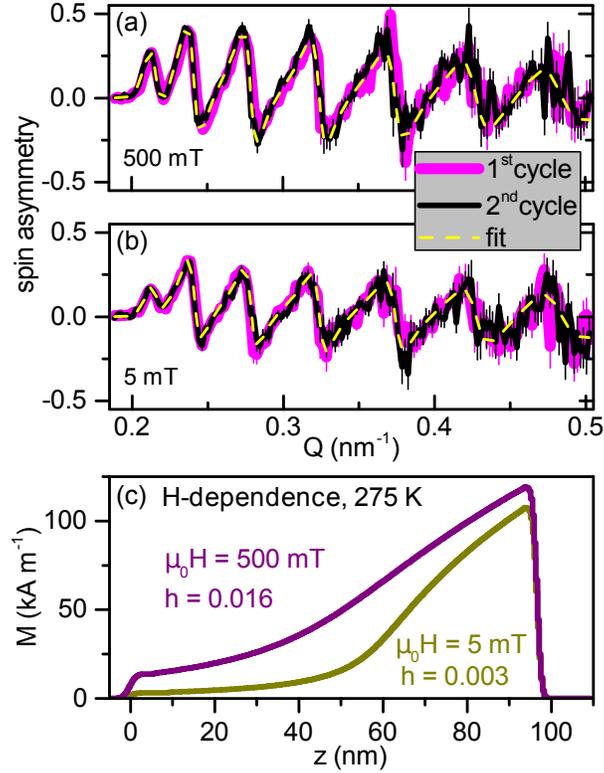}
\caption{\label{fig4} (a-b) Fitted PNR data plotted as spin asymmetry show clear reversible behavior, as well as a strong field dependent amplitude.  (c) Magnetic profiles corresponding to the fits in (a-b).  Error bars correspond to $\pm$ 1 standard deviation.}
\end{figure}

This work shows that a smoothly graded 100 nm Ni$_{x(z)}$Cu$_{1-x(z)}$ alloy film exhibits a ferromagnetic phase transition similar to what one would expect from a continuum of uncoupled ferromagnetic layers with distinct Curie temperatures.  Interestingly, interlayer exchange coupling plays a negligible role in the spatial evolution of the ferromagnetic phase transition.  This is particularly surprising because temperature dependent exchange coupling can in some cases be strong enough to fully reverse ferromagnets.\cite{Zhipan_PRL_2006}  As such, we demonstrate continuous control of the displacement of a quasi-phase boundary between weakly and strongly magnetically ordered regions in a thin film structure.  At temperatures between the minimum and maximum local Curie temperature, the magnetization on the weakly magnetized side of the boundary can be enhanced significantly and reversibly with application of a relatively modest applied field.  These functionalities were designed to be most active near room temperature, but the physics applies to any temperature range that may be desired.  The extension to magnetic systems with technological relevance, such as FePt, is obvious.  Since the exchange length in these high anisotropy systems is only a few nanometers, it should be possible to engineer thin FePt films that exhibit similar behavior dictated by gradients in composition.\cite{Dumas_PRB_2011}  Further, since the volume of ferromagnetically ordered spins changes with temperature in these structures, one can envision novel application in thermomagnetic sensors and switches.\cite{Kravets_PRB_2012}  For instance, a symmetrically graded structure with the paramagnetic region in the center could show complex temperature dependent magnetic coupling, since the thickness of the paramagnetic interlayer region will be temperature dependent.  In addition, rationally designed composition profiles could be employed to achieve specifically desired temperature and field dependencies.  Such an approach could be used to tailor the magnetocaloric response of a medium to realize advances in applications such as magnetic refrigeration.\cite{Sandeman_Scripta_2012,Miller_JVST_2014}   

\begin{acknowledgments}
Work at RIT was supported by NSF-CAREER $\#$1522927.  Work at nanoGUNE was supported by the Spanish Government (Project No. MAT2012-36844).
\end{acknowledgments}

\begin{thebibliography}{23}%
\makeatletter
\providecommand \@ifxundefined [1]{%
 \@ifx{#1\undefined}
}%
\providecommand \@ifnum [1]{%
 \ifnum #1\expandafter \@firstoftwo
 \else \expandafter \@secondoftwo
 \fi
}%
\providecommand \@ifx [1]{%
 \ifx #1\expandafter \@firstoftwo
 \else \expandafter \@secondoftwo
 \fi
}%
\providecommand \natexlab [1]{#1}%
\providecommand \enquote  [1]{``#1''}%
\providecommand \bibnamefont  [1]{#1}%
\providecommand \bibfnamefont [1]{#1}%
\providecommand \citenamefont [1]{#1}%
\providecommand \href@noop [0]{\@secondoftwo}%
\providecommand \href [0]{\begingroup \@sanitize@url \@href}%
\providecommand \@href[1]{\@@startlink{#1}\@@href}%
\providecommand \@@href[1]{\endgroup#1\@@endlink}%
\providecommand \@sanitize@url [0]{\catcode `\\12\catcode `\$12\catcode
  `\&12\catcode `\#12\catcode `\^12\catcode `\_12\catcode `\%12\relax}%
\providecommand \@@startlink[1]{}%
\providecommand \@@endlink[0]{}%
\providecommand \url  [0]{\begingroup\@sanitize@url \@url }%
\providecommand \@url [1]{\endgroup\@href {#1}{\urlprefix }}%
\providecommand \urlprefix  [0]{URL }%
\providecommand \Eprint [0]{\href }%
\providecommand \doibase [0]{http://dx.doi.org/}%
\providecommand \selectlanguage [0]{\@gobble}%
\providecommand \bibinfo  [0]{\@secondoftwo}%
\providecommand \bibfield  [0]{\@secondoftwo}%
\providecommand \translation [1]{[#1]}%
\providecommand \BibitemOpen [0]{}%
\providecommand \bibitemStop [0]{}%
\providecommand \bibitemNoStop [0]{.\EOS\space}%
\providecommand \EOS [0]{\spacefactor3000\relax}%
\providecommand \BibitemShut  [1]{\csname bibitem#1\endcsname}%
\let\auto@bib@innerbib\@empty
\bibitem [{\citenamefont {Ramos}\ \emph {et~al.}(1990)\citenamefont {Ramos},
  \citenamefont {Lederman}, \citenamefont {King},\ and\ \citenamefont
  {Jaccarino}}]{Ramos_PRL_1990}%
  \BibitemOpen
  \bibfield  {author} {\bibinfo {author} {\bibfnamefont {C.~A.}\ \bibnamefont
  {Ramos}}, \bibinfo {author} {\bibfnamefont {D.}~\bibnamefont {Lederman}},
  \bibinfo {author} {\bibfnamefont {A.~R.}\ \bibnamefont {King}}, \ and\
  \bibinfo {author} {\bibfnamefont {V.}~\bibnamefont {Jaccarino}},\ }\href@noop
  {} {\bibfield  {journal} {\bibinfo  {journal} {Phys. Rev. Lett.}\ }\textbf
  {\bibinfo {volume} {65}},\ \bibinfo {pages} {2913} (\bibinfo {year}
  {1990})}\BibitemShut {NoStop}%
\bibitem [{\citenamefont {Wang}\ and\ \citenamefont
  {Mills}(1992)}]{Wang_PRB_1992}%
  \BibitemOpen
  \bibfield  {author} {\bibinfo {author} {\bibfnamefont {R.~W.}\ \bibnamefont
  {Wang}}\ and\ \bibinfo {author} {\bibfnamefont {D.~L.}\ \bibnamefont
  {Mills}},\ }\href@noop {} {\bibfield  {journal} {\bibinfo  {journal} {Phys.
  Rev. B}\ }\textbf {\bibinfo {volume} {46}},\ \bibinfo {pages} {11681}
  (\bibinfo {year} {1992})}\BibitemShut {NoStop}%
\bibitem [{\citenamefont {Marcellini}\ \emph {et~al.}(2009)\citenamefont
  {Marcellini}, \citenamefont {Parnaste}, \citenamefont {Hjorvarsson},\ and\
  \citenamefont {Wolff}}]{Marcellini_PRB_2009}%
  \BibitemOpen
  \bibfield  {author} {\bibinfo {author} {\bibfnamefont {M.}~\bibnamefont
  {Marcellini}}, \bibinfo {author} {\bibfnamefont {M.}~\bibnamefont
  {Parnaste}}, \bibinfo {author} {\bibfnamefont {B.}~\bibnamefont
  {Hjorvarsson}}, \ and\ \bibinfo {author} {\bibfnamefont {M.}~\bibnamefont
  {Wolff}},\ }\href@noop {} {\bibfield  {journal} {\bibinfo  {journal} {Phys.
  Rev. B}\ }\textbf {\bibinfo {volume} {79}},\ \bibinfo {pages} {144426}
  (\bibinfo {year} {2009})}\BibitemShut {NoStop}%
\bibitem [{\citenamefont {Suess}(2006)}]{Suess_APL_2006}%
  \BibitemOpen
  \bibfield  {author} {\bibinfo {author} {\bibfnamefont {D.}~\bibnamefont
  {Suess}},\ }\href@noop {} {\bibfield  {journal} {\bibinfo  {journal} {Appl.
  Phys. Lett.}\ }\textbf {\bibinfo {volume} {89}},\ \bibinfo {pages} {113105}
  (\bibinfo {year} {2006})}\BibitemShut {NoStop}%
\bibitem [{\citenamefont {Berger}\ \emph {et~al.}(2008)\citenamefont {Berger},
  \citenamefont {Supper}, \citenamefont {Ikeda}, \citenamefont {Lengsfield},
  \citenamefont {Moser},\ and\ \citenamefont {Fullerton}}]{Berger_APL_2008}%
  \BibitemOpen
  \bibfield  {author} {\bibinfo {author} {\bibfnamefont {A.}~\bibnamefont
  {Berger}}, \bibinfo {author} {\bibfnamefont {N.}~\bibnamefont {Supper}},
  \bibinfo {author} {\bibfnamefont {Y.}~\bibnamefont {Ikeda}}, \bibinfo
  {author} {\bibfnamefont {B.}~\bibnamefont {Lengsfield}}, \bibinfo {author}
  {\bibfnamefont {A.}~\bibnamefont {Moser}}, \ and\ \bibinfo {author}
  {\bibfnamefont {E.~E.}\ \bibnamefont {Fullerton}},\ }\href@noop {} {\bibfield
   {journal} {\bibinfo  {journal} {Appl. Phys. Lett.}\ }\textbf {\bibinfo
  {volume} {93}},\ \bibinfo {pages} {122502} (\bibinfo {year}
  {2008})}\BibitemShut {NoStop}%
\bibitem [{\citenamefont {Berger}\ \emph {et~al.}(2010)\citenamefont {Berger},
  \citenamefont {Fullerton}, \citenamefont {Van~Do},\ and\ \citenamefont
  {Supper}}]{berger2010magnetic}%
  \BibitemOpen
  \bibfield  {author} {\bibinfo {author} {\bibfnamefont {A.}~\bibnamefont
  {Berger}}, \bibinfo {author} {\bibfnamefont {E.}~\bibnamefont {Fullerton}},
  \bibinfo {author} {\bibfnamefont {H.}~\bibnamefont {Van~Do}}, \ and\ \bibinfo
  {author} {\bibfnamefont {N.}~\bibnamefont {Supper}},\ }\href
  {http://www.google.com/patents/US7687157} {} (\bibinfo {year} {2010}),\
  \bibinfo {note} {uS Patent 7,687,157}\BibitemShut {NoStop}%
\bibitem [{\citenamefont {Kirby}\ \emph {et~al.}(2010)\citenamefont {Kirby},
  \citenamefont {Davies}, \citenamefont {Liu}, \citenamefont {Watson},
  \citenamefont {Zimanyi}, \citenamefont {Shull}, \citenamefont {Kienzle},\
  and\ \citenamefont {Borchers}}]{Kirby_PRB_2010}%
  \BibitemOpen
  \bibfield  {author} {\bibinfo {author} {\bibfnamefont {B.~J.}\ \bibnamefont
  {Kirby}}, \bibinfo {author} {\bibfnamefont {J.~E.}\ \bibnamefont {Davies}},
  \bibinfo {author} {\bibfnamefont {K.}~\bibnamefont {Liu}}, \bibinfo {author}
  {\bibfnamefont {S.~M.}\ \bibnamefont {Watson}}, \bibinfo {author}
  {\bibfnamefont {G.~T.}\ \bibnamefont {Zimanyi}}, \bibinfo {author}
  {\bibfnamefont {R.~D.}\ \bibnamefont {Shull}}, \bibinfo {author}
  {\bibfnamefont {P.~A.}\ \bibnamefont {Kienzle}}, \ and\ \bibinfo {author}
  {\bibfnamefont {J.~A.}\ \bibnamefont {Borchers}},\ }\href {\doibase
  10.1103/PhysRevB.81.100405} {\bibfield  {journal} {\bibinfo  {journal} {Phys.
  Rev. B}\ }\textbf {\bibinfo {volume} {81}},\ \bibinfo {pages} {100405}
  (\bibinfo {year} {2010})}\BibitemShut {NoStop}%
\bibitem [{\citenamefont {Kirby}\ \emph {et~al.}(2015)\citenamefont {Kirby},
  \citenamefont {Greene}, \citenamefont {Maranville}, \citenamefont {Davies},\
  and\ \citenamefont {Liu}}]{Kirby_JAP_2015}%
  \BibitemOpen
  \bibfield  {author} {\bibinfo {author} {\bibfnamefont {B.~J.}\ \bibnamefont
  {Kirby}}, \bibinfo {author} {\bibfnamefont {P.~K.}\ \bibnamefont {Greene}},
  \bibinfo {author} {\bibfnamefont {B.~B.}\ \bibnamefont {Maranville}},
  \bibinfo {author} {\bibfnamefont {J.~E.}\ \bibnamefont {Davies}}, \ and\
  \bibinfo {author} {\bibfnamefont {K.}~\bibnamefont {Liu}},\ }\href@noop {}
  {\bibfield  {journal} {\bibinfo  {journal} {J. Appl. Phys.}\ }\textbf
  {\bibinfo {volume} {117}},\ \bibinfo {pages} {063905} (\bibinfo {year}
  {2015})}\BibitemShut {NoStop}%
\bibitem [{\citenamefont {LeGra{\"e}t}\ \emph {et~al.}(2015)\citenamefont
  {LeGra{\"e}t}, \citenamefont {Charlton}, \citenamefont {McLaren},
  \citenamefont {Loving}, \citenamefont {Morley}, \citenamefont {Kinane},
  \citenamefont {Brydson}, \citenamefont {Lewis}, \citenamefont {Langridge},\
  and\ \citenamefont {Marrows}}]{LeGraet_APLmat_2015}%
  \BibitemOpen
  \bibfield  {author} {\bibinfo {author} {\bibfnamefont {C.}~\bibnamefont
  {LeGra{\"e}t}}, \bibinfo {author} {\bibfnamefont {T.~R.}\ \bibnamefont
  {Charlton}}, \bibinfo {author} {\bibfnamefont {M.}~\bibnamefont {McLaren}},
  \bibinfo {author} {\bibfnamefont {M.}~\bibnamefont {Loving}}, \bibinfo
  {author} {\bibfnamefont {S.~A.}\ \bibnamefont {Morley}}, \bibinfo {author}
  {\bibfnamefont {C.~J.}\ \bibnamefont {Kinane}}, \bibinfo {author}
  {\bibfnamefont {R.~M.~D.}\ \bibnamefont {Brydson}}, \bibinfo {author}
  {\bibfnamefont {L.~H.}\ \bibnamefont {Lewis}}, \bibinfo {author}
  {\bibfnamefont {S.}~\bibnamefont {Langridge}}, \ and\ \bibinfo {author}
  {\bibfnamefont {C.~H.}\ \bibnamefont {Marrows}},\ }\href@noop {} {\bibfield
  {journal} {\bibinfo  {journal} {APL Materials}\ }\textbf {\bibinfo {volume}
  {3}},\ \bibinfo {pages} {041802} (\bibinfo {year} {2015})}\BibitemShut
  {NoStop}%
\bibitem [{\citenamefont {Skomski}\ and\ \citenamefont
  {Sellmyer}(2000)}]{Skomski_JAP_2000}%
  \BibitemOpen
  \bibfield  {author} {\bibinfo {author} {\bibfnamefont {R.}~\bibnamefont
  {Skomski}}\ and\ \bibinfo {author} {\bibfnamefont {D.~J.}\ \bibnamefont
  {Sellmyer}},\ }\href@noop {} {\bibfield  {journal} {\bibinfo  {journal} {J.
  Appl. Phys.}\ }\textbf {\bibinfo {volume} {87}},\ \bibinfo {pages} {4756}
  (\bibinfo {year} {2000})}\BibitemShut {NoStop}%
\bibitem [{\citenamefont {Ahern}\ \emph {et~al.}(1958)\citenamefont {Ahern},
  \citenamefont {Martin},\ and\ \citenamefont {Sucksmith}}]{Ahern_PRS_1958}%
  \BibitemOpen
  \bibfield  {author} {\bibinfo {author} {\bibfnamefont {S.~A.}\ \bibnamefont
  {Ahern}}, \bibinfo {author} {\bibfnamefont {M.~J.~C.}\ \bibnamefont
  {Martin}}, \ and\ \bibinfo {author} {\bibfnamefont {W.~A.}\ \bibnamefont
  {Sucksmith}},\ }\href@noop {} {\bibfield  {journal} {\bibinfo  {journal}
  {Proc. Roy. Soc. A}\ }\textbf {\bibinfo {volume} {248}},\ \bibinfo {pages}
  {145} (\bibinfo {year} {1958})}\BibitemShut {NoStop}%
\bibitem [{\citenamefont {Kravets}\ \emph {et~al.}(2012)\citenamefont
  {Kravets}, \citenamefont {Timoshevskii}, \citenamefont {Yanchitsky},
  \citenamefont {Bergmann}, \citenamefont {Buhler}, \citenamefont {Andersson},\
  and\ \citenamefont {Korenivski}}]{Kravets_PRB_2012}%
  \BibitemOpen
  \bibfield  {author} {\bibinfo {author} {\bibfnamefont {A.~F.}\ \bibnamefont
  {Kravets}}, \bibinfo {author} {\bibfnamefont {A.~N.}\ \bibnamefont
  {Timoshevskii}}, \bibinfo {author} {\bibfnamefont {B.~Z.}\ \bibnamefont
  {Yanchitsky}}, \bibinfo {author} {\bibfnamefont {M.~A.}\ \bibnamefont
  {Bergmann}}, \bibinfo {author} {\bibfnamefont {J.}~\bibnamefont {Buhler}},
  \bibinfo {author} {\bibfnamefont {S.}~\bibnamefont {Andersson}}, \ and\
  \bibinfo {author} {\bibfnamefont {V.}~\bibnamefont {Korenivski}},\
  }\href@noop {} {\bibfield  {journal} {\bibinfo  {journal} {Phys. Rev. B}\
  }\textbf {\bibinfo {volume} {86}},\ \bibinfo {pages} {214413} (\bibinfo
  {year} {2012})}\BibitemShut {NoStop}%
\bibitem [{\citenamefont {Berger}\ and\ \citenamefont
  {Fullerton}(1997)}]{Berger_JMMM_1997}%
  \BibitemOpen
  \bibfield  {author} {\bibinfo {author} {\bibfnamefont {A.}~\bibnamefont
  {Berger}}\ and\ \bibinfo {author} {\bibfnamefont {E.~E.}\ \bibnamefont
  {Fullerton}},\ }\href@noop {} {\bibfield  {journal} {\bibinfo  {journal} {J.
  Magn. Magn. Mater.}\ }\textbf {\bibinfo {volume} {165}},\ \bibinfo {pages}
  {471} (\bibinfo {year} {1997})}\BibitemShut {NoStop}%
\bibitem [{\citenamefont {Riego}\ and\ \citenamefont
  {Berger}(2015)}]{Riego_PRE_2015}%
  \BibitemOpen
  \bibfield  {author} {\bibinfo {author} {\bibfnamefont {P.}~\bibnamefont
  {Riego}}\ and\ \bibinfo {author} {\bibfnamefont {A.}~\bibnamefont {Berger}},\
  }\href@noop {} {\bibfield  {journal} {\bibinfo  {journal} {Phys. Rev. E}\
  }\textbf {\bibinfo {volume} {91}},\ \bibinfo {pages} {062141} (\bibinfo
  {year} {2015})}\BibitemShut {NoStop}%
\bibitem [{\citenamefont {Binder}\ and\ \citenamefont
  {Hohenberg}(1974)}]{Binder_PRB_1974}%
  \BibitemOpen
  \bibfield  {author} {\bibinfo {author} {\bibfnamefont {K.}~\bibnamefont
  {Binder}}\ and\ \bibinfo {author} {\bibfnamefont {P.~C.}\ \bibnamefont
  {Hohenberg}},\ }\href@noop {} {\bibfield  {journal} {\bibinfo  {journal}
  {Phys. Rev. B}\ }\textbf {\bibinfo {volume} {9}},\ \bibinfo {pages} {2194}
  (\bibinfo {year} {1974})}\BibitemShut {NoStop}%
\bibitem [{sup()}]{supplemental}%
  \BibitemOpen
  \href@noop {} {}\bibinfo {note} {See Supplemental Material at [URL will be
  inserted by publisher] for details}\BibitemShut {NoStop}%
\bibitem [{\citenamefont {Majkrzak}\ \emph {et~al.}(2005)\citenamefont
  {Majkrzak}, \citenamefont {O'Donovan},\ and\ \citenamefont
  {Berk}}]{Chuck_book}%
  \BibitemOpen
  \bibfield  {author} {\bibinfo {author} {\bibfnamefont {C.~F.}\ \bibnamefont
  {Majkrzak}}, \bibinfo {author} {\bibfnamefont {K.~V.}\ \bibnamefont
  {O'Donovan}}, \ and\ \bibinfo {author} {\bibfnamefont {N.~F.}\ \bibnamefont
  {Berk}},\ }in\ \href@noop {} {\emph {\bibinfo {booktitle} {Neutron Scattering
  from Magnetic Materials}}},\ \bibinfo {editor} {edited by\ \bibinfo {editor}
  {\bibfnamefont {T.}~\bibnamefont {Chatterji}}}\ (\bibinfo  {publisher}
  {Elsevier Science},\ \bibinfo {address} {New York},\ \bibinfo {year}
  {2005})\BibitemShut {NoStop}%
\bibitem [{Note1()}]{Note1}%
  \BibitemOpen
  \bibinfo {note} {We note that lifting this constraint, and allowing $z_{C}$
  to vary freely at each temperature does not significantly improve the fit
  quality, and yields $T$-dependent profiles very similar to those shown in
  Fig. 4(b).}\BibitemShut {Stop}%
\bibitem [{Note2()}]{Note2}%
  \BibitemOpen
  \bibinfo {note} {For the modeling, $T_{C}'$ corresponds to zero $m$ in zero
  $H$, while $T_{C}$ for the boundary condition samples was defined as the
  minimum in ${\begingroup dm\endgroup \over dT}$ in finite $H$. Thus the
  absolute disagreement among the values is largely semantic.}\BibitemShut
  {Stop}%
\bibitem [{\citenamefont {Li}\ \emph {et~al.}(2006)\citenamefont {Li},
  \citenamefont {Eisenmenger}, \citenamefont {Miller},\ and\ \citenamefont
  {Schuller}}]{Zhipan_PRL_2006}%
  \BibitemOpen
  \bibfield  {author} {\bibinfo {author} {\bibfnamefont {Z.-P.}\ \bibnamefont
  {Li}}, \bibinfo {author} {\bibfnamefont {J.}~\bibnamefont {Eisenmenger}},
  \bibinfo {author} {\bibfnamefont {C.~W.}\ \bibnamefont {Miller}}, \ and\
  \bibinfo {author} {\bibfnamefont {I.~K.}\ \bibnamefont {Schuller}},\ }\href
  {\doibase 10.1103/PhysRevLett.96.137201} {\bibfield  {journal} {\bibinfo
  {journal} {Phys. Rev. Lett.}\ }\textbf {\bibinfo {volume} {96}},\ \bibinfo
  {pages} {137201} (\bibinfo {year} {2006})}\BibitemShut {NoStop}%
\bibitem [{\citenamefont {Dumas}\ \emph {et~al.}(2011)\citenamefont {Dumas},
  \citenamefont {Fang}, \citenamefont {Kirby}, \citenamefont {Zha},
  \citenamefont {Bonanni}, \citenamefont {Nogues},\ and\ \citenamefont
  {{\AA}kerman}}]{Dumas_PRB_2011}%
  \BibitemOpen
  \bibfield  {author} {\bibinfo {author} {\bibfnamefont {R.~K.}\ \bibnamefont
  {Dumas}}, \bibinfo {author} {\bibfnamefont {Y.}~\bibnamefont {Fang}},
  \bibinfo {author} {\bibfnamefont {B.~J.}\ \bibnamefont {Kirby}}, \bibinfo
  {author} {\bibfnamefont {C.}~\bibnamefont {Zha}}, \bibinfo {author}
  {\bibfnamefont {V.}~\bibnamefont {Bonanni}}, \bibinfo {author} {\bibfnamefont
  {J.}~\bibnamefont {Nogues}}, \ and\ \bibinfo {author} {\bibfnamefont
  {J.}~\bibnamefont {{\AA}kerman}},\ }\href {\doibase
  10.1103/PhysRevB.84.054434} {\bibfield  {journal} {\bibinfo  {journal} {Phys.
  Rev. B}\ }\textbf {\bibinfo {volume} {84}},\ \bibinfo {pages} {054434}
  (\bibinfo {year} {2011})}\BibitemShut {NoStop}%
\bibitem [{\citenamefont {Sandeman}(2012)}]{Sandeman_Scripta_2012}%
  \BibitemOpen
  \bibfield  {author} {\bibinfo {author} {\bibfnamefont {K.~G.}\ \bibnamefont
  {Sandeman}},\ }\href@noop {} {\bibfield  {journal} {\bibinfo  {journal}
  {Scripta Materialia}\ }\textbf {\bibinfo {volume} {67}},\ \bibinfo {pages}
  {566} (\bibinfo {year} {2012})}\BibitemShut {NoStop}%
\bibitem [{\citenamefont {Miller}\ \emph {et~al.}(2014)\citenamefont {Miller},
  \citenamefont {Belyea},\ and\ \citenamefont {Kirby}}]{Miller_JVST_2014}%
  \BibitemOpen
  \bibfield  {author} {\bibinfo {author} {\bibfnamefont {C.~W.}\ \bibnamefont
  {Miller}}, \bibinfo {author} {\bibfnamefont {D.~D.}\ \bibnamefont {Belyea}},
  \ and\ \bibinfo {author} {\bibfnamefont {B.~J.}\ \bibnamefont {Kirby}},\
  }\href@noop {} {\bibfield  {journal} {\bibinfo  {journal} {Journal of Vacuum
  Science and Technology A}\ }\textbf {\bibinfo {volume} {32}},\ \bibinfo
  {pages} {040802} (\bibinfo {year} {2014})}\BibitemShut {NoStop}%
\end{thebibliography}
\providecommand{\noopsort}[1]{}\providecommand{\singleletter}[1]{#1}%

\end{document}